\documentclass[10pt]{iopart}
\usepackage{graphicx}
\usepackage{subfigure}
\usepackage{rotating}
\usepackage{moreverb}
\usepackage{url}
\usepackage{epsf}
\usepackage{multirow}
\usepackage{morefloats}
\usepackage{threeparttable}
\usepackage{bm}
\usepackage{booktabs}
\usepackage{braket}
\usepackage[dvipsnames,table,xcdraw]{xcolor}
\usepackage{iopams}
\usepackage{cite}

\begin{document}

\ioptwocol
%\begin{abstract}
%We present a computational approach which is tailored for reducing the complexity of the description of extended systems at the density functional theory level.
%We define a recipe for generating a set of localized basis functions which are optimized either for the accurate description of pristine, bulk like Wannier functions, or for the \emph{in situ} treatment of deformations induced by defective constituents such as boundaries or impurities. Our method enables one to identify the regions of an extended system which require dedicated optimization of the Kohn-Sham degrees of freedom, and provides the user with a reliable estimation of the errors -- if any -- induced by the locality of the approach.
%Such a method facilitates on the one hand an effective reduction of the computational degrees of freedom needed to simulate systems at the nanoscale, while in turn providing a description that can be straightforwardly put in relation to effective models, like tight binding Hamiltonians.
%We present our methodology with SiC nanotube-like cages as a test bed.
%Nonetheless, the wavelet-based method employed in this paper makes possible calculation of systems with different dimensionalities, including slabs and fully periodic systems.
%\end{abstract}

%\maketitle

\twocolumn[
\begin{@twocolumnfalse}
%Adaptive Localized Orbitals as a Tool for Studying Defect Locality in Linear-Scaling Density Functional Theory Simulations
%\title{Embedded fragment approach for large scale defect systems I}
%Edge Effects
\title{Pseudo-Fragment Approach for Extended Systems Derived from Linear-Scaling DFT}%: Application to SiC Nanotubes}

\author{Laura E.\ Ratcliff}
\ead{laura.ratcliff08@imperial.ac.uk}
\address{Department of Materials, Imperial College London, London SW7 2AZ, UK}
\address{Argonne Leadership Computing Facility, Argonne National Laboratory, Illinois 60439, USA}

\author{Luigi Genovese}
\address{Univ.\ Grenoble Alpes, CEA, INAC-MEM, L\_Sim, F-38000, Grenoble, France}

\date{\today}
%\vspace{10pt}
%\begin{indented}
%\item[]\today
%\end{indented}
%\maketitle
\begin{abstract}
We present a computational approach which is tailored for reducing the complexity of the description of extended systems at the density functional theory level.
We define a recipe for generating a set of localized basis functions which are optimized either for the accurate description of pristine, bulk like Wannier functions, or for the \emph{in situ} treatment of deformations induced by defective constituents such as boundaries or impurities. Our method enables one to identify the regions of an extended system which require dedicated optimization of the Kohn-Sham degrees of freedom, and provides the user with a reliable estimation of the errors -- if any -- induced by the locality of the approach.
Such a method facilitates on the one hand an effective reduction of the computational degrees of freedom needed to simulate systems at the nanoscale, while in turn providing a description that can be straightforwardly put in relation to effective models, like tight binding Hamiltonians.
We present our methodology with SiC nanotube-like cages as a test bed.
Nonetheless, the wavelet-based method employed in this paper makes possible calculation of systems with different dimensionalities, including slabs and fully periodic systems.
\end{abstract}
\end{@twocolumnfalse}
]

\section{Introduction}

Density-functional theory (DFT)~\cite{hohenberg42,kohn43} is arguably the most popular method for electronic structure calculations, and is routinely applied to a range of materials and properties for systems containing up to a few hundred atoms.  In recent years, the development of linear scaling (LS) approaches~\cite{Goedecker1999,Bowler2012}, which overcome the computational limits of traditional cubic scaling implementations, has enabled the treatment of thousands 
of atoms from first principles. Such LS methods open up the possibility not just of treating larger systems, but also new types of materials and calculations~\cite{Ratcliff2017}.

One popular route to LS involves the creation of a minimal set of localized support functions (SFs), which are expressed in an underlying systematic basis set and optimized \emph{in situ} to reflect their chemical environment. Importantly, this approach, which has been adopted in \textsc{onetep}~\cite{Skylaris2005}, \textsc{Conquest}~\cite{Bowler2010} and \textsc{BigDFT}~\cite{Mohr2014,Mohr2015}, preserves the same high level of accuracy as standard cubic scaling approaches employing systematic basis sets, such as plane waves or wavelets.
This brings accurate DFT calculations of very large systems into reach, although the prefactor associated with the SF optimization can be high.

In tandem with the development of LS-DFT, various fragment and embedding methods have also been developed to reduce the computational cost of treating large systems using DFT, in some cases in conjunction with other quantum mechanical methods.
Such approaches have in common the division of a target system into two or more subsystems and are generally applied in one of two ways.  In the case of fragment methods, this is primarily as a `divide and conquer' approach to reducing the cost of treating a system at a single level of theory,
along the same lines as Yang's LS-DFT approach of the same name~\cite{Yang1991}.
In other words, all fragments are treated on equal footing, e.g.~\cite{Fedorov2008,Fedorov2009,Fedorov2017,Genova2017}.  Embedding methods, on the other hand, are generally used as a means of implementing a multi-scale approach, wherein one fragment (the active region) is treated using a higher level of theory, while the remainder of the system (the environment) is treated with a computationally less expensive method, e.g.~\cite{Manby2012,Fornace2015,Cheng2017,Cheng2017b,Culpitt2017,Ding2017,Libisch2017,Chulhai2018,Hegely2018}.  The active region is typically treated using a higher order quantum chemistry method, hybrid functional DFT and/or a larger basis set, while the environment is treated with for example semi-local DFT and/or a smaller basis set.

The accuracy of an embedding calculation depends on both the choice of method for treating the interactions between subsystems and the way in which the system is divided.
For molecular systems, the fragments are usually the constituent molecules of a supramolecular system, or a larger molecule and surrounding solvent molecules.
Along these lines, we and others have recently implemented a molecular fragment approach~\cite{Ratcliff2015a} in the wavelet-based \textsc{BigDFT} code~\cite{Genovese2008}.  This approach takes advantage of similarity between molecules with similar local environments, by using SFs generated for isolated `template' fragments, in this case molecules, and using them as a fixed basis for a large system of many such molecules.
The fragmentation is thus at the level of the basis, with the fragments fully interacting in the final calculation.
The approach is well suited to materials which consist of clearly separated fragments.
We have therefore recently formalized such a fragmentation concept~\cite{Mohr2017}, showing that a suitable inspection of density matrix related quantities of a full LS calculation may be used to determine whether or not a fragment can be seen as a separable `moiety' of the whole system.

In extended systems, however, there is often no clear fragmentation scheme, so that the assignment of an atom to a particular subsystem is somewhat arbitrary.
Furthermore, depending on the method used to account for the interactions, both the accuracy and computational cost can be highly sensitive to the chosen fragmentation.
Nonetheless, to our knowledge no scheme exists to define a sensible fragmentation \emph{a priori}.  
As such, a means of predicting whether or not a given fragmentation choice is reasonable would be highly useful.
To this end, in the following we explore the advantages of using optimized SFs from LS-DFT both to set up a fragment-like approach and as a means of gaining insight into the suitability of a given fragmentation of an extended system.

Given that the SFs in LS-DFT are optimized to reflect their local environment, if two SFs associated with different atoms are the same, this implies that their environments are also the same.  For an infinitely repeating system this should be the case, to within some level of numerical noise coming from the representation in a systematic but nonetheless incomplete basis.  However, where there is a defect or deformation, SFs close to the defect would differ significantly from those far away.  If one could quantify the similarity between two SFs, it would thus be possible to assess the physical extent of the defect, i.e.\ the distance over which it has a significant effect on the SFs and thus the electronic structure. 

To show how such a quantity might be used in practice, we introduce in this work a pseudo-fragment method based on LS-DFT which is able to treat extended systems containing some degree of repetition, e.g.\ a periodic system comprising many repeat units.  Such an approach might easily be defined if, instead of taking an isolated template fragment, one generates the SF basis by embedding the pseudo-fragment in a representative environment.
As with the molecular fragment approach, this can result in considerable computational savings.  More interestingly it also provides the opportunity to explore in detail the errors introduced by different levels of approximation.

Using the pseudo-fragment approach, we show how the similarity between SFs located on atoms with similar chemical environments relates to the accuracy of a given pseudo-fragment setup.
That is, the SF similarity provides a means of determining to what extent a given fragmentation of an extended system is reasonable, i.e.\ whether or not an atom should best be associated with an active region where further SF optimization is needed or an environment region where the template SFs are already able to accurately represent the electronic structure.  Furthermore, such a measure can also be used to give a reliable estimation of the errors induced by the imposed pseudo-fragmentation.
For the sake of conciseness in this work we focus on the case study of a one-dimensional system, in this case a nanotube, however we stress that the approach could easily be adapted to other extended systems.
Similarly, although we have implemented the approach in \textsc{BigDFT}, it could also be implemented in other LS-DFT codes which generate optimized SFs.

The layout of this manuscript is as follows.  In Section~\ref{sec:theory} we recap the approach to LS-DFT implemented in \textsc{BigDFT} and explain the key concepts of the molecular fragment approach.
In Section~\ref{sec:sicnt} we then present the embedded pseudo-fragment approach as applied to a SiC nanotube (NT).  We consider both an infinitely repeating NT and a finite NT, quantifying the similarity between SFs on equivalent atoms and demonstrating the applicability of template SFs generated in a short NT for a much longer NT.
In Section~\ref{sec:wahba} we then consider the effect of introducing deformations to the system, first in the form of random noise in the atomic coordinates, and then by considering finite nanotubes of varying lengths.  Finally, in Section~\ref{sec:conclusion}, we present a summary and outlook on future applications of our method.

\section{\label{sec:theory}Background Theory: Molecular Fragments in BigDFT}

\subsection{Linear Scaling BigDFT}

Thanks to the nearsightedness principle, in the $\mathcal O(N)$ formalism implemented in the BigDFT code,
we assume that the density matrix of the system $\hat F$ can be defined from a set of localized SFs $|\phi_\alpha\rangle$ as follows:
\begin{equation}
 \hat F =\sum_{\alpha,\beta} |\phi_\alpha\rangle K^{\alpha\beta}\langle \phi_\beta| \;,
 \label{eq:density_matrix_in_support_function_basis}
\end{equation}
with a SF overlap matrix $S_{\alpha\beta} = \braket{\phi_\alpha | \phi_\beta}$,
which can be chosen to have a unit diagonal and where $K^{\alpha\beta}$ is the so-called density kernel.
In the LS-BigDFT approach, the SFs are expressed in a Daubechies wavelet basis and are defined such as to optimize the target function
\begin{equation}
 \Omega = \mathrm{tr}\left( \hat F \hat H_c\right)
\end{equation}
where the operator $\hat H_c = \hat H_{\mathrm{KS}}[\rho] + \hat V_c$ is the sum of the density-dependent Kohn-Sham (KS) Hamiltonian
plus a confining operator $\hat V_c$ such that
\begin{eqnarray}
 \bra{\phi_\alpha} \hat V_c \ket{\phi_\beta} &=& \delta_{\alpha\beta} \bra{\phi_\alpha} \hat V_c^\alpha \ket{\phi_\alpha}\;, \\
 \quad V_c^\alpha(\mathbf r) &=& c_\alpha | \mathbf r - \mathbf R_\alpha |^4 \;,
\end{eqnarray}
which has the purpose of keeping the SFs confined in their localization regions,
centered in the position $\mathbf R_\alpha$, while
reducing the KS band structure energy. Usually $\mathbf R_\alpha$ coincides with the position
$\mathbf R_a$ of the atom $a$ where $\phi_\alpha$ is initially centered at the beginning of the
SCF optimization procedure.
To some extent this enables one to associate $\alpha$ to a particular atom $a$.
The coeffcient $c_\alpha$ is dynamically adjusted during the basis set optimization procedure.
For a molecular calculation, we obtain in this way a \emph{minimal} set of molecular orbitals that,
by construction, exactly represent the occupied KS orbitals, and also has a non-zero projection to the
unoccupied orbitals subspace.
In an extended system, although the SFs resulting from LS-BigDFT are entirely numerical and are therefore not constrained to
any particular form, they generally retain some resemblance to atomic orbitals,
and are thus referred to as e.g.\ $s$-like SFs.
For a given SF basis, the density kernel is obtained using a choice of methods -- in the following we use the Fermi operator expansion (FOE) approach~\cite{Mohr2014,Mohr2015}.

\subsection{Molecular Fragment Approach}

When simulating a system that is made from a collection of various molecules
it would appear reasonable to associate to each of these a `fragment'
whose electronic structure might be represented by a subset of the SFs of the entire system.
Such an approach has been already presented in \cite{Ratcliff2015a,Ratcliff2015b} and further formalized
in \cite{Mohr2017}, where a recipe has been provided to
identify such fragments out of the density matrix of the full system.
The use of such a fragment approach substantially reduces the computational cost, and has been applied to the calculation of electronic charge transport parameters in a disordered supramolecular morphology~\cite{Ratcliff2015b}.

The molecular fragment approach in \textsc{BigDFT} is conceptually straightforward, and can be summarized in three steps:
\begin{enumerate}
\item \textbf{template calculation:} generate the SF basis $\left\{ \phi_\alpha^\mathfrak F\right\}$
for each unique fragment type $\mathfrak F$ by performing a full LS calculation for an isolated fragment,
optimizing both the SFs and density kernel.
\item \textbf{SF replication:} take the SFs from the isolated template calculations and replicate them for the full system of interest.
\item \textbf{full calculation:} perform a LS calculation for the full system using the replicated SFs as a fixed basis, optimizing only the density kernel.
\end{enumerate}
Both the first and last steps use the standard machinery of a LS calculation.
The second step can be easily automated, however,
it is complicated by the need to account for a suitable set of rototranslation transformations.
Namely, given the range of potential orientations and positions of fragments in a particular target system,
it is essential to have a method which is able to rototranslate the template SFs from their template coordinates
to the correct location and orientation in the full system of interest.

\subsubsection{Rototranslation procedure}
For the case where there is no internal deformation of a given fragment, i.e.\ the transformation from template to system fragment is a rigid rotation, one can easily find the appropriate transformation. 
As such, we therefore apply the rotation matrix $\mathcal R^{T\rightarrow S}$ which minimizes the following cost function $J$
\begin{equation}\label{eq:wahba}
J\left(\mathcal{R}^{T\rightarrow S}\right)=\frac{1}{2}\sum_{a=1}^{N}||\mathbf{R}_a^S-\sum_{b=1}^{N}\mathcal{R}^{T\rightarrow S}_{ab}\mathbf{R}_a^T||^2\;,
\end{equation}
where $N$ is the number of atoms in the fragment and $\mathbf{R}_a^{T(S)}$ are the coordinates of the template (system) fragment.
The identification of the optimal transformation from such a cost function is a well known problem~\cite{Wahba1965,Kabsch1978}, and may easily be found using a simple singular value decomposition based approach~\cite{Markley1988}.
Such a transformation may then be combined with the translation $\mathcal T^{T\rightarrow S}$ which
is straightforwardly defined from the respective centre of mass shifts.
Having found the appropriate rototranslation,
we then use a wavelet-based interpolation scheme to perform the identified transformation,
as explained in \cite{Ratcliff2015a}.
The computational cost of this interpolation step is kept low,
as is the error introduced by this rototranslation.

To summarize, let us assume that a system is made of a collection of fragments $\left\{ \mathfrak F_i^{(\ell)}\right\}$,
with the index $\ell$ labelling the fragment type and the index $i$ running over all the fragments of the same type.
A set of fragment templates $\left\{ \mathfrak F_0^{(\ell)}\right\}$ can then be chosen and the associated SFs
$\left\{\phi_\alpha^{(\ell)}\right\}$ extracted as per step (\textrm i) above.
The rototranslation procedure guarantees that the calculation of step (\textrm{iii}) employs the basis set defined as
$\left\{\mathcal{RT}^{\ell_0\rightarrow \ell_i} \phi_\alpha^{(\ell)}\right\}$ where the rototranslation
$\mathcal{RT}^{\ell_0\rightarrow \ell_i}$ connects the template fragment of type $\ell$ with the $i$-th moiety of the same type.
Such a rototranslation minimizes by construction the cost function $J$ of Eq.~(\ref{eq:wahba}).

The approach described above is performed at the level of the SFs rather than the density operator,
so that the fragments are still fully able to interact with each other \emph{via} the full system density
without the need to account for overlapping fragments.
Nonetheless, the approach is designed for the creation of a molecular basis, which relies on two key assumptions.
First, that the fragments are separable, i.e.\ that the density matrix can be reasonably approximated by a `block-like'
matrix, each block being associated to a fragment, and that the interaction between neighbouring
fragments is weak enough to not significantly impact on the form of the SFs with respect to the template basis set.
Second, that the fragments are close to rigid, i.e.\ that the internal deformations between different fragment instances remain small.  For the latter, the cost function of Eq.~\ref{eq:wahba} may be used to quantify the level of deformation.  For large values of $J$, the internal deformations are too high and the fragment approximation breaks down.

\section{\label{sec:sicnt}Embedded Fragment Approach: Case Study of a SiC Nanotube}

The approach described so far is well-suited for molecular systems,
or more precisely for systems where the fundamental constituents have molecular character.
The key question is therefore whether it is possible to define an equivalent approach for extended systems, wherein a template basis is generated from a small representative system.  In this section we focus on responding to this question.

We take the case study of a (quasi-)one dimensional system, namely a SiC nanotube.  We begin with a pristine SiCNT in periodic boundary conditions, for which the structure is depicted in Fig.~\ref{fig:sic_posinp_per}.
We first assess the extent to which the SFs follow our expectations of similarity along the length of the NT (i.e.\ equivalent SFs for equivalent environments), and then demonstrate how to take advantage of such repetition in practice.
In order to quantify the error resulting from the strict localization of the SFs, we also perform initial calculations using the cubic scaling approach of \textsc{BigDFT}, wherein the KS wavefunctions are calculated directly without any additional localization constraints.
To this end, we take a long NT containing 14 repeat units, comprising 392 atoms in total;
this is both short enough to be accessible to the cubic scaling
approach of \textsc{BigDFT} and long enough to avoid the need for $k$-point sampling.
Computational details are given in~\ref{app:comp}.

\subsection{Assessing Basis Similarity: Onsite Overlap Matrix}\label{sec:onsite_overlap}

\begin{figure}
\centering
\subfigure[Supercell atomic structure.]{\includegraphics[width=0.5\textwidth]{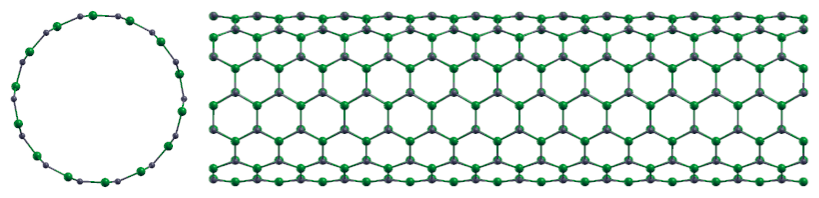}\label{fig:sic_posinp_per}}
\subfigure[Select optimized support functions.]{\includegraphics[height=0.49\textwidth,angle=-90]{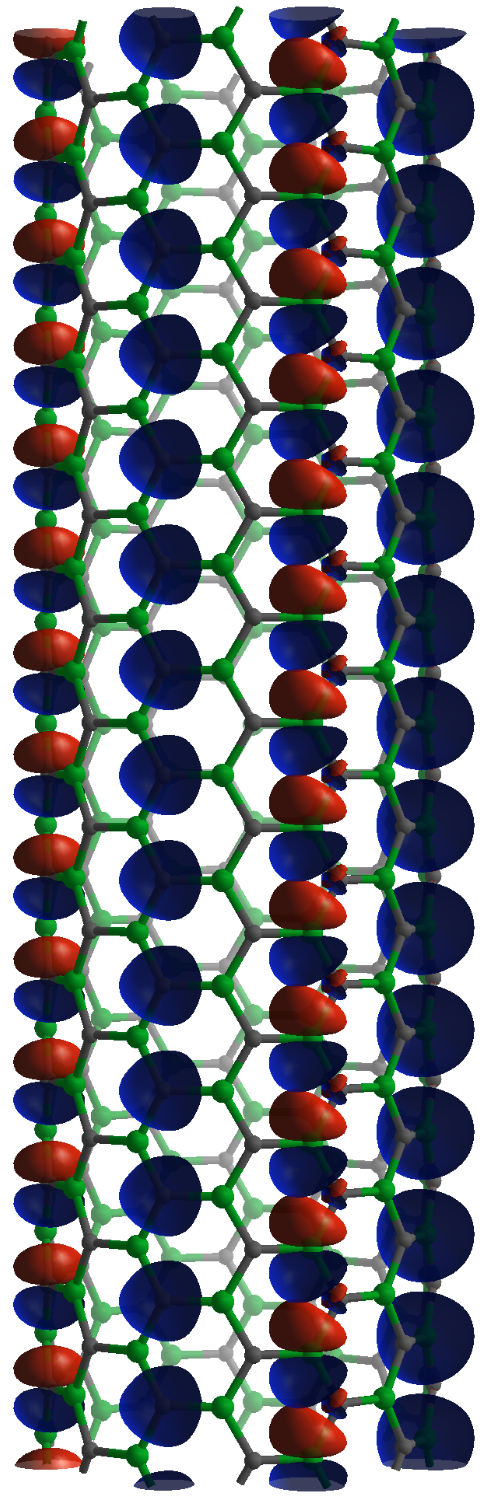}\label{fig:sic_tmb_per}}
\subfigure[Onsite overlap.]{\includegraphics[scale=0.31]{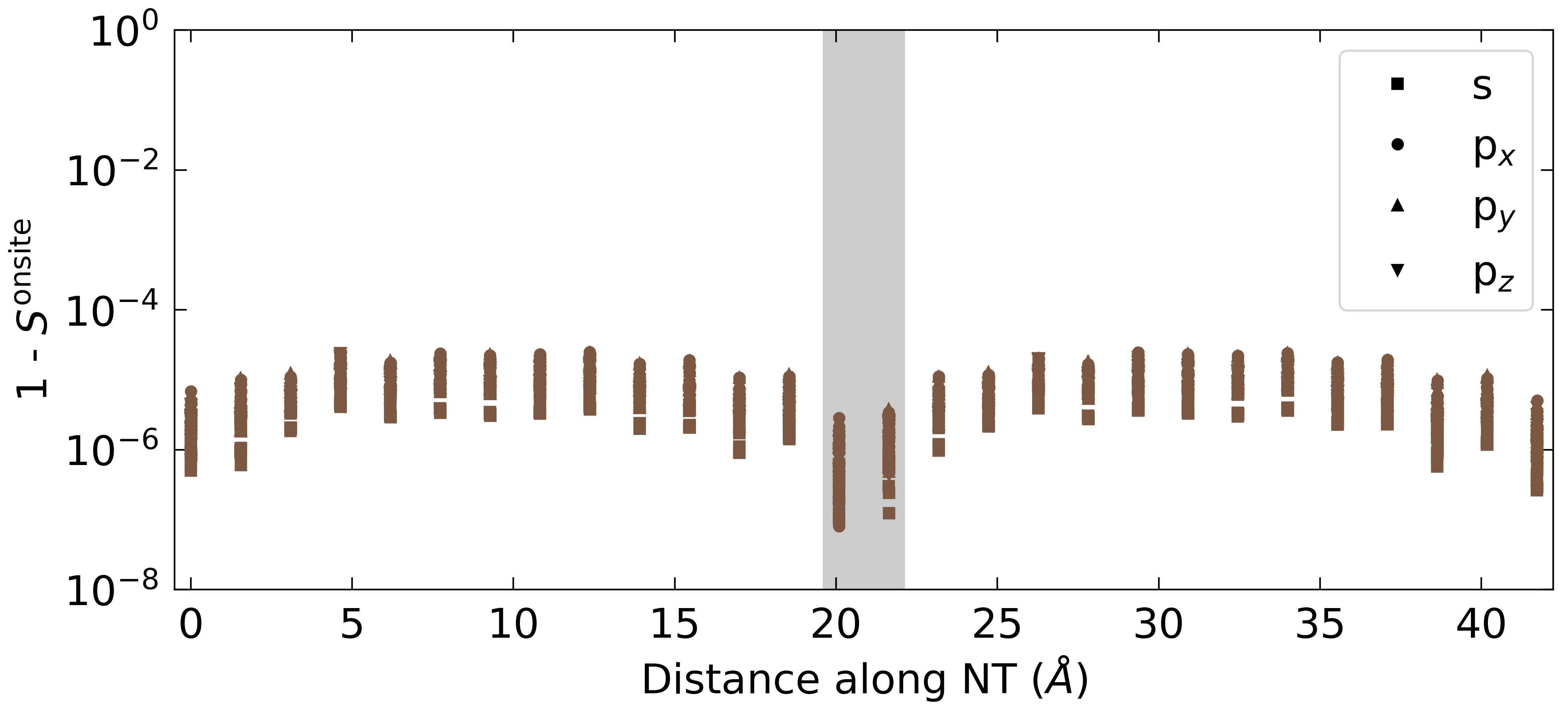}\label{fig:sic_oo_per}}
\caption{Panel \ref{fig:sic_posinp_per}: pristine SiCNT treated with periodic boundary conditions, comprising 14 NT units (392 atoms).
Si (C) are atoms depicted in green (grey).
Visualization (panel \ref{fig:sic_tmb_per}) of select optimized $s$ and $p_z$ type SFs
and corresponding $S^{\mathrm{onsite}}$ values (panel \ref{fig:sic_oo_per}).
Onsite overlap values have been calculated between each SF and that of the equivalent
type (i.e.\ between $s$ and $s$-type functions) and the equivalent atom in the centre of the NT
(i.e.\ between atom $i$ of fragment $j$ and atom $i$ of the central fragment),
where the central region is shaded in grey.
The values in the central region determine the error associated with the numerical noise of the reformatting procedure. 
}\label{fig:sic_per}
\end{figure}

Before considering how to employ a fragment-like approach for the SiCNT, we first pose some questions on the nature of the optimized SFs.
In this case we do not have the notion of `moiety', as the system cannot be partitioned into distinctly separable fragments.
However we expect that for a pristine SiCNT, the SFs in equivalent locations along the length of the NT should be quasi-identical, with only small differences due to numerical noise arising from the egg-box effect.  Select optimized SFs resulting from a standard LS calculation are depicted in Fig.~\ref{fig:sic_tmb_per}, where it can be seen that the selected SFs are indeed qualitatively the same.  However, it would be useful to have a more reliable measure than visual inspection, and furthermore to have a \emph{quantitative} measure of the degree of similarity.

To this end, we define a quantity which we refer to as the onsite overlap matrix, $S^{\textrm{onsite}}$.
This is defined as
\begin{equation}
 S^{\textrm{onsite}}_{\alpha \beta} \equiv
 \braket{\mathcal{RT}^{\mathbf R_\alpha \rightarrow \mathbf R_0}\phi_\alpha^{(\ell)} | \mathcal{RT}^{\mathbf R_\beta \rightarrow \mathbf R_0}\phi_\beta^{(\ell)}} \;,
\end{equation}
i.e.\ as the overlap between two SFs (which may or may not originally be centred on the same atom) if they were to be centred on the same site of centre $\mathbf R_0$.
Where $S_{\alpha\beta}^{\textrm{onsite}}=1$, this implies SFs $\alpha$ and $\beta$ are identical, presumably due to having both the same character ($s$, $p_x$ etc.) and the same local environment, irrespective of whether there is a spatial overlap.
The onsite overlap is thus independent of the physical location of the SFs and is instead purely a measure of the degree of similarity between two SFs, and correspondingly the chemical environment of the atoms with which they are associated.

We calculate $S^{\textrm{onsite}}$ by first choosing a localization region associated with an arbitrary atom of
position $\mathbf R_0$, then using the interpolation scheme discussed above to reformat all SFs into this region.
The rotations -- if needed -- should be chosen such as to properly align the local environment of the atoms associated to the indices
$\alpha$ and $\beta$.
Following this, the full onsite overlap matrix can be calculated using the same machinery used to calculate the standard (spatial) SF overlap matrix.  This process introduces some numerical noise, whose size depends on the employed grid spacing.
We therefore do not expect to see exact agreement between two SFs in equivalent environments,
however this discrepancy should be much smaller than the signal in which we are interested.
Such an indicator should therefore help in answering the question of to what extent the presence of a defect (e.g.\ a boundary or an impurity) at a given distance might result in differences between SFs.

Returning to the specific example of the SiCNT, we have taken a ring of atoms in the centre of the NT as a reference,
and calculated onsite overlap matrix elements between SFs of the same type centred on equivalent atoms along the NT,
i.e.\ with the same $x$ and $y$ coordinates but different $z$ coordinates. The various elements of the
system are therefore simply translated with respect to each other.
For symmetry reasons, we expect that such a choice of the reference SFs would provide a unit value of $S^\textrm{onsite}$.
The results of $1-S^\textrm{onsite}$ are plotted in Fig.~\ref{fig:sic_oo_per}.
The values in the central region (highlighted in grey) should be zero, since they correspond to the overlap between a
SF and itself; the values in this region therefore give an estimate of the noise level associated with the wavelet
grid spacing.
Although there is some variation along the NT, the values remain small throughout,
in line with our observations that the SFs in different rings of the NT are indeed equivalent.
Such a measure therefore helps us in \emph{quantifying} to what extent the value of $S^\textrm{onsite}$
might deviate from unity before the SFs can no longer be considered to be equivalent.
In some sense, it may be ascribed to a \emph{calibration} procedure of the fundamental constituents of the
system's density matrix.

\subsection{(Pseudo) Fragments in a Periodic System}

Given the above, one might imagine that, rather than optimizing the SFs for the entire NT, one could take the SFs from a single slice, and replicate them across the NT.
Assuming that the fluctuations in the onsite overlap matrix are not meaningful (i.e.\ purely noise), this approach should have little impact on the accuracy of the calculation.
We note that although there are other possible divisions of a SiCNT into fragments, we choose to take the simple definition of one complete (covalently bonded) ring of the NT (28 atoms) for convenience.
Unlike in a molecular system, such a fragment is not in any sense separable from the full system,
so we use the term pseudo-fragment (PFrag) to distinguish it from its molecular counterpart.

The question becomes, how does one perform the equivalent of a molecular template calculation, i.e.\ how can one generate the SFs for a PFrag which are appropriate for an extended system?  It is clear that optimizing the SFs for an isolated ring would be a poor approach.
Instead, following the principles of nearsightedness, 
we propose to perform a template calculation wherein the PFrag is \emph{embedded} in a representative environment.
Given a sufficient number of nearest neighbours (i.e.\ neighbouring slices), this should result in a SF basis which is of a similar quality as that which has been directly generated in the full system.

\subsection{Pseudo-Fragment Workflow}
\begin{figure}
\centering
\subfigure[Pseudo-Fragment setup, with one PFrag type generated in a six unit periodic NT template.]{\includegraphics[scale=0.22]{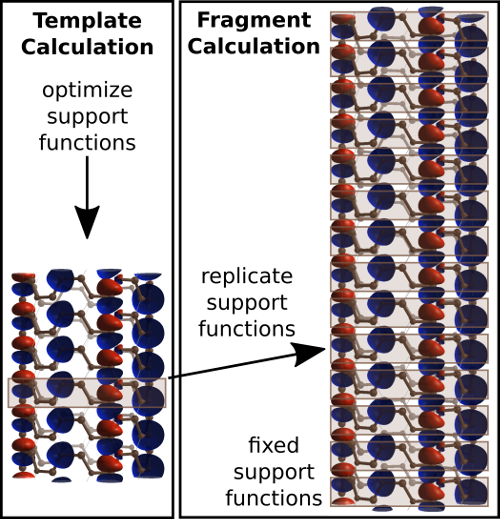}\label{fig:sic_schem_per}}
\subfigure[Density of states.]{\includegraphics[scale=0.3]{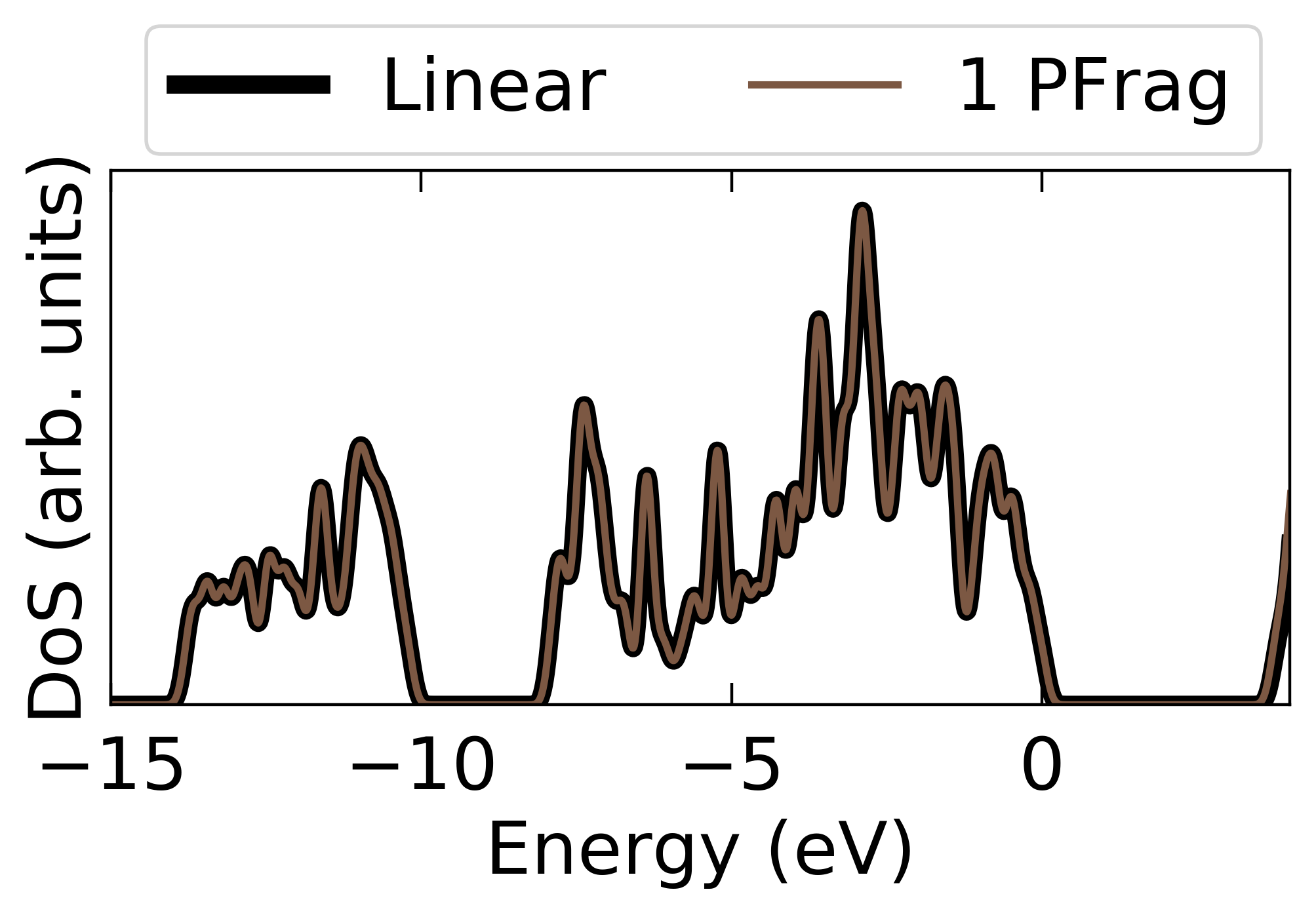}\label{fig:sic_per_dos}}
\subfigure[Averaged electron densities.]{\includegraphics[scale=0.3]{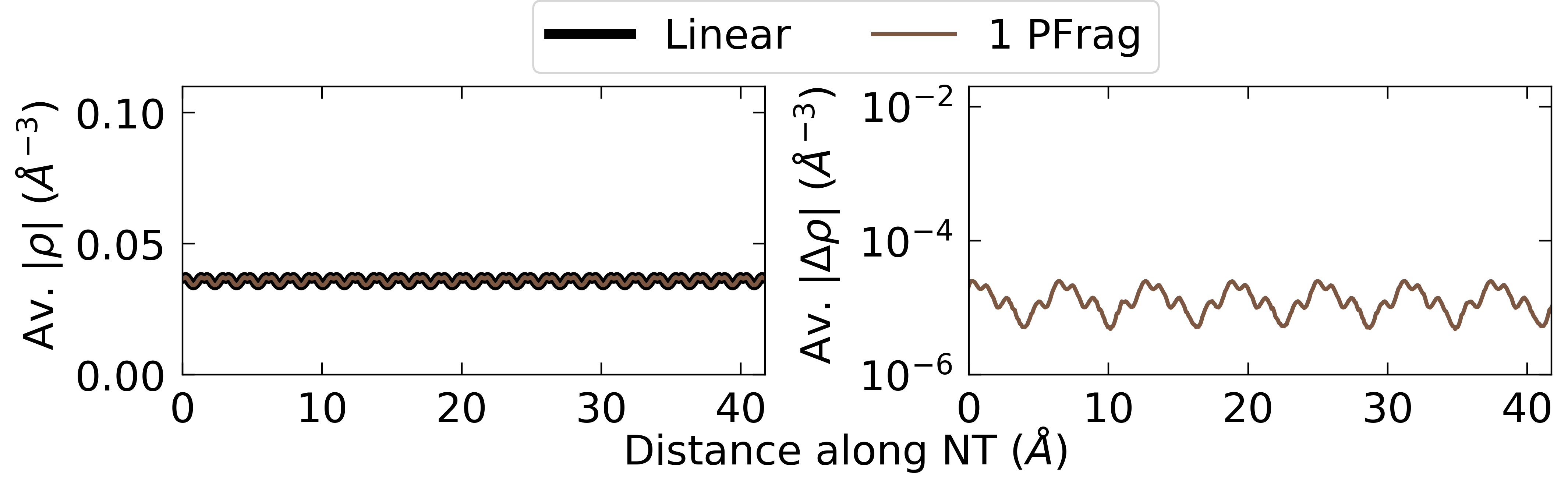}\label{fig:sic_per_rho}}
\caption{Panel \ref{fig:sic_schem_per}: schematic illustrating the PFrag setup for the periodic SiCNT of Fig.~\ref{fig:sic_per}. Si atoms are depicted as larger spheres.
Panel \ref{fig:sic_per_dos}: DoS for the PFrag and LS approaches.
Panel \ref{fig:sic_per_rho}: electron densities ($\rho$) averaged over the $xy$-plane [left] and their corresponding errors ($\Delta \rho$) relative
to the LS reference [right]. }
\end{figure}

The workflow for the SiCNT is thus as depicted in Fig.~\ref{fig:sic_schem_per}.
The template SFs are generated for a shorter SiCNT periodic supercell, in this case containing 6 ring units
(168 atoms). 
The SFs associated with a given ring are then replicated for the 14 unit calculation,
while the SFs from other five rings are discarded (although they could also be taken into account by averaging, see~\ref{app:mfi}).
Aside from the need for an embedded template calculation, the calculation follows along similar lines as a molecular fragment calculation, with the SFs remaining fixed for the full system, requiring only the (self-consistent) optimization of the density kernel.
The only exception is that building the density kernel from embedded fragments is less straightforward than in the isolated case.  Here and in the following we choose to build the initial guess for the density kernel directly from the density kernel of the template calculation(s), as discussed in~\ref{app:kernel}.

\begin{table*}
  \caption{\label{tab:sic_energies}
Total energies $E$, errors in energy compared to the LS reference (i.e.\ $\Delta E = E - E^{\textrm{linear}}$), the band gap $\Delta_{\mathrm{gap}}$, and average and maximum values of the cost function $J$ for different calculation setups.  For the pseudo-fragment calculations, also specified is whether or not the template SFs were generated for a periodic SiCNT, finite SiCNT, or a combination.  The predicted error is derived from the noisy periodic calculations (see Section~\ref{sec:noise}) and is based purely on the average value of $J$.}
\begin{tabular}{ll c rr c r c rr c r}
\toprule
&  && $E$ & $\Delta E$ && $\Delta_{\mathrm{gap}}$ && $J_{\mathrm{Av.}}$ & $J_{\mathrm{Max.}}$ && Pred. $\Delta E$\\
& Template && eV/atom & meV/atom && eV && \AA$^2$ & \AA$^2$ && meV/atom\\

\cline{1-2} \cline{4-5} \cline{7-7} \cline{9-10} \cline{12-12}\\[-2.5ex]
\textbf{Periodic NT}\\
Linear & -&& -130.793 & 0.0 && 3.79 && - & - && - \\
1 PFrag & Per. && -130.789 & 3.3 && 3.74 && 0.000 & 0.000 && 3.4\\
\cline{1-2} \cline{4-5} \cline{7-7} \cline{9-10} \cline{12-12} \\[-2.5ex]
\textbf{Finite NT}\\
Linear & - && -130.719 & 0.0 && 3.63 && - & - && -  \\
1 PFrag & Per. &&-129.438 &	1280.6 && 1.45 && 4.3473 & 30.2885 && 30821.7\\
4 PFrags & Fin.  && -129.770 & 948.9 && 3.08 &&  0.1712 & 0.2760 && 1216.8 \\
6 PFrags & Fin. && -130.675 & 44.0 && 3.59 && 0.0157 & 0.0309 && 114.6\\
8 PFrags & Fin.  && -130.712 & 7.0 && 3.62 && 0.0010 & 0.0021 && 10.6\\
10 PFrags & Fin.  && -130.714 & 4.8 && 3.62 && 0.0003 & 0.0012 && 5.0\\
7 PFrags & 6 Fin.\ + 1 Per.  && -130.710 & 9.2 && 3.62 && 0.0007 & 0.0018 && 8.2\\
\bottomrule
\end{tabular}
\end{table*}

The total energies and band gaps resulting from the LS and PFrag setups
are given in Table~\ref{tab:sic_energies}, where it can be seen that the PFrag approach
results in a negligible error in the total energy -- around 3~meV/atom.
The band gap is also in good agreement.  It should be noted that the band gap is significantly overestimated compared to the cubic scaling approach, by more than 1~eV.
This is to be expected, since the SF basis is explicitly optimized to represent the occupied KS states.  In situations were one requires access to one or two unoccupied states, one might explicitly add them into the SF optimization procedure, in which case it is necessary to use the direct minimization approach for density kernel optimization~\cite{Mohr2014}.  For a larger number of states, it would be necessary to use an approach such as the one used in the \textsc{onetep} code and optimize a second set of SFs specifically to represent the unoccupied states~\cite{Ratcliff2011}.  For the purposes of this work, however, we are interested only in demonstrating the ability to reproduce the band gap of the optimized SF basis, which we can see is the case to within $0.05$~eV.

The density of states (DoS) for the two methods are depicted in Fig.~\ref{fig:sic_per_dos}, where the KS energies were obtained from a one-off diagonalization in the SF basis at the end of the LS and PFrag calculations.  The curves for the two approaches are indistinguishable at the given level of smearing.
Although not depicted here, we note that aside from the aforementioned discrepancy for the unoccupied states, the linear and cubic scaling DoS are also in excellent agreement, demonstrating that the PFrag approach is indeed able to reproduce the correct occupied DoS.

Finally, the average electronic densities along the axis of the SiCNT are plotted in Fig.~\ref{fig:sic_per_rho}.  As with the DoS, the average density of the PFrag approach is indistinguishable from the LS calculation.  Looking also at the average differences along the NT with respect to the reference LS density, it is clear that the error for the PFrag approach is very small compared to the absolute values.  For reference, the average density difference between the linear and cubic scaling calculations is of the order of $10^{-4}$.
In short, by all of the above measures, one can conclude that the PFrag approach is indeed able to accurately reproduce the electronic structure of the full SiCNT.

Once the PFrag basis is generated, such an approach provides in principle a set of \emph{pre-optimized}
basis functions, whose utility is twofold: on one hand, they offer a basis set that might
be employed \emph{as-is} for the simulation of large pristine systems,
enabling considerable computational savings compared to optimizing the SFs for the full system,
as well as a set of \emph{fixed} degrees of freedom on which `second principles' Hamiltonians (like for example tight binding models) can be designed.
On the other hand, more interestingly, the pre-optimized basis set equipped with the
calculation of $S^\textrm{onsite}$ provide a \emph{gauge}
to measure the actual extension of defective regions, i.e.\ the region for which
the pristine SFs cannot be used to obtain high levels of precision.
We will see in the following an example of this procedure.

\begin{figure}
 \centering
\subfigure[Atomic structure.]{\includegraphics[width=0.5\textwidth]{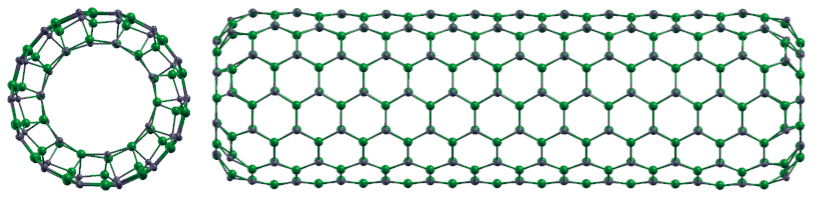}\label{fig:sic_posinp_free}}
\subfigure[Select optimized support functions.]{\includegraphics[height=0.49\textwidth,angle=-90]{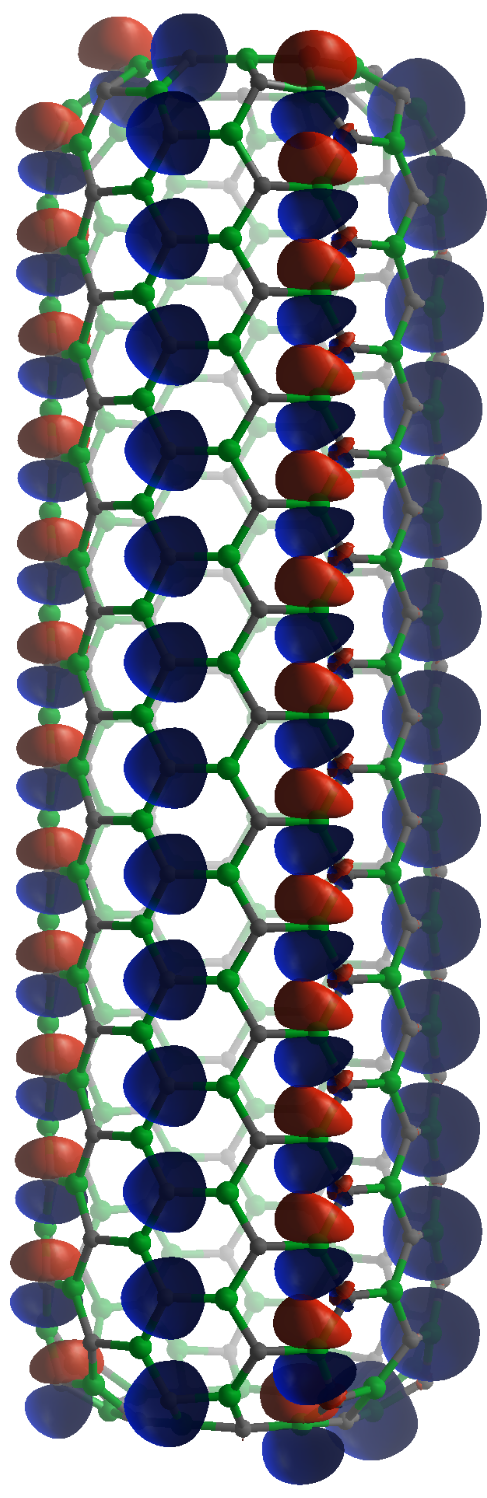}\label{fig:sic_tmb_free}}
\subfigure[Onsite overlap.]{\includegraphics[scale=0.31]{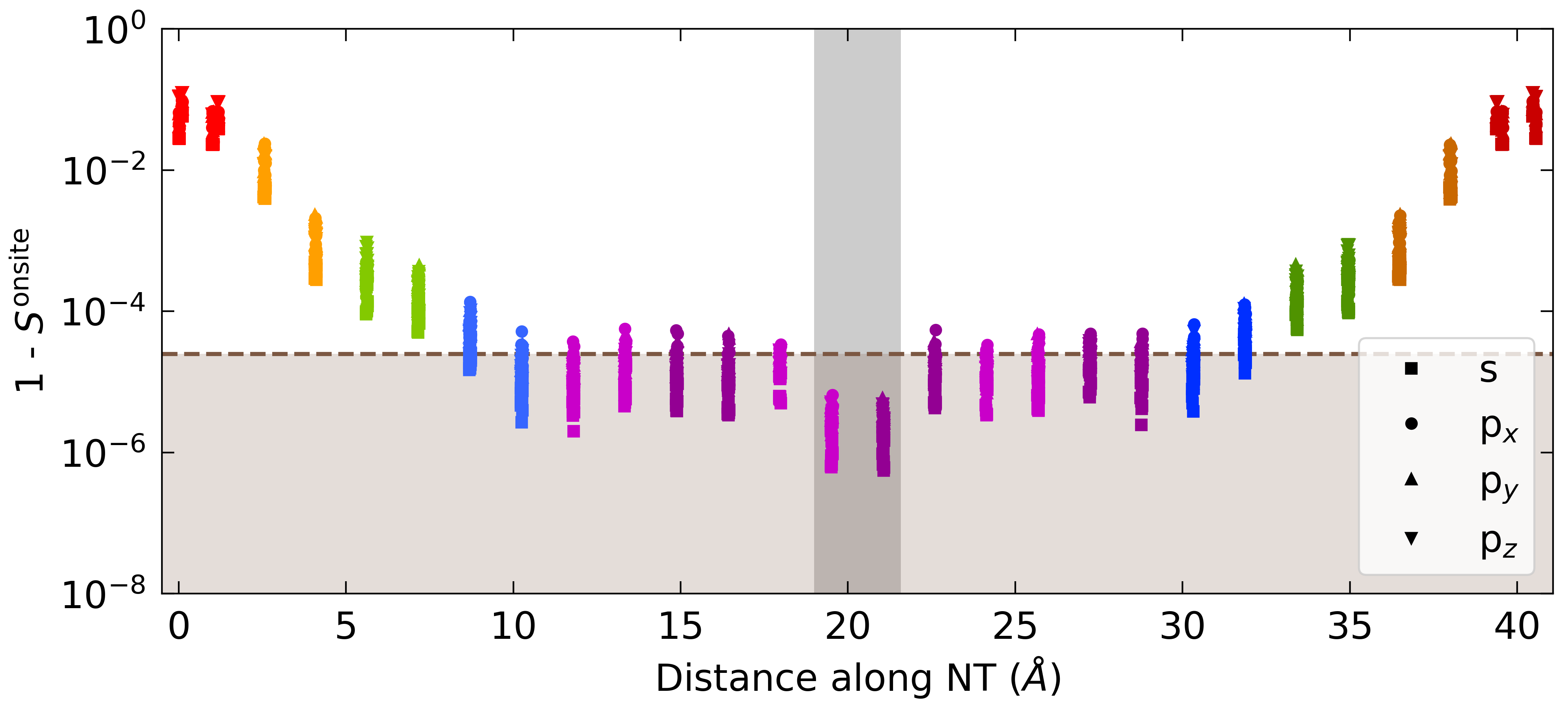}\label{fig:sic_oo_free}}
\subfigure[Correlation between onsite overlap and error in energy.]{\includegraphics[scale=0.31]{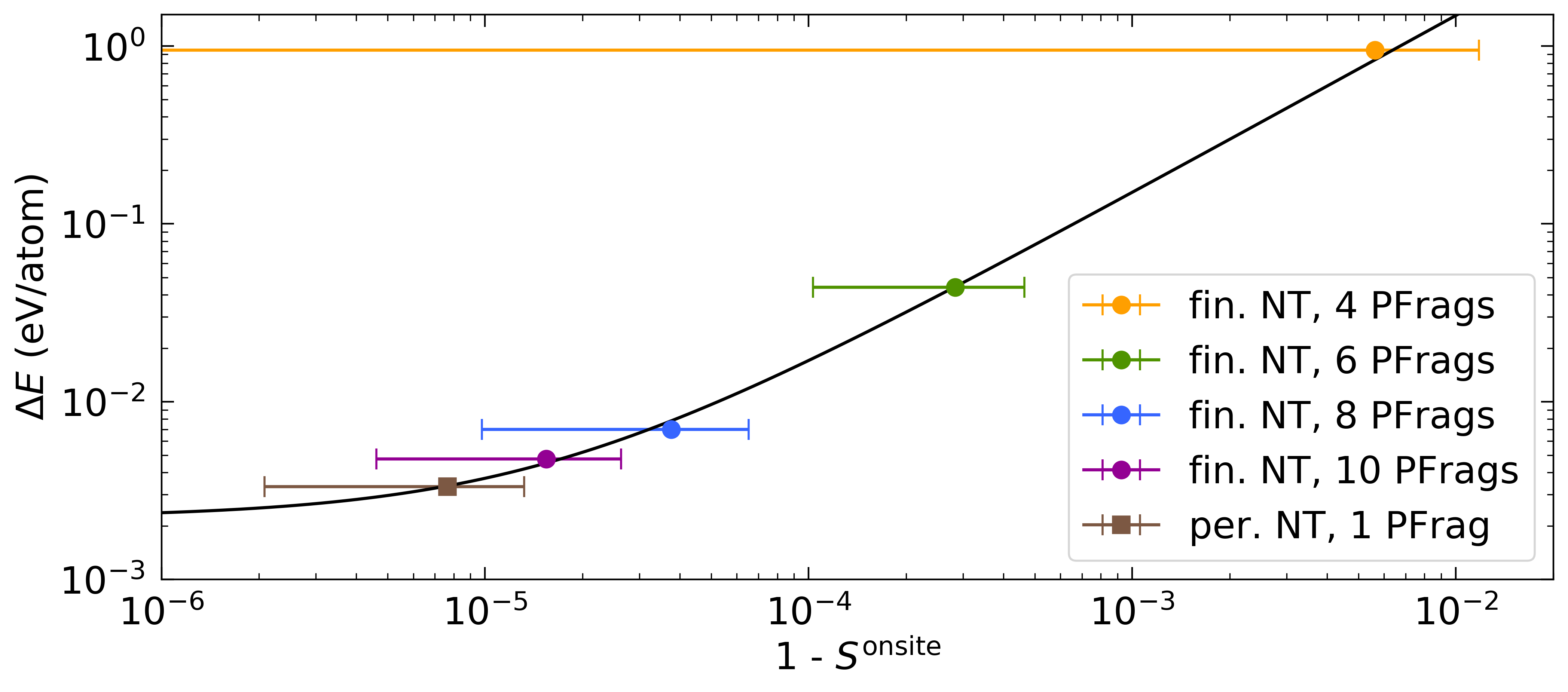}\label{fig:sic_oo_v_e}}
\caption{Panel \ref{fig:sic_posinp_free}: finite SiCNT treated with open boundary conditions, comprising 14 NT units (392 atoms).
Si (C) are atoms depicted in green (grey).
Visualization (panel \ref{fig:sic_tmb_free}) of select optimized $s$ and $p_z$ type SFs and corresponding $S^{\mathrm{onsite}}$ values (panel \ref{fig:sic_oo_free}), calculated following the prescription described in Fig.~\ref{fig:sic_per}.  The dashed line represents the maximum value for $S^{\mathrm{onsite}}$ for the periodic NT and thus the shaded area corresponds to differences due to numerical noise. Panel 3(d): correlation between $S^{\mathrm{onsite}}$ and error in energy ($\Delta E = E - E^{\mathrm{linear}}$). The $S^{\mathrm{onsite}}$ values are the average values for a given PFrag type as calculated in the full length NT, i.e.\ those plotted in panel \ref{fig:sic_oo_free}.  The error bars indicate the standard deviation. We note there is a relatively large variation in onsite overlap matrix values within PFrag types, particularly for the PFrag second in from the edge.  This is primarily due to the choice of PFrag -- if we took instead a single ring of (unbonded) C atoms, i.e.\ a PFrag of half the size, this variation would be much smaller.
}\label{fig:sic_free}
\end{figure}

\begin{figure*}
\centering
\subfigure[One PFrag type generated in a six unit periodic NT template.]{\includegraphics[scale=0.22]{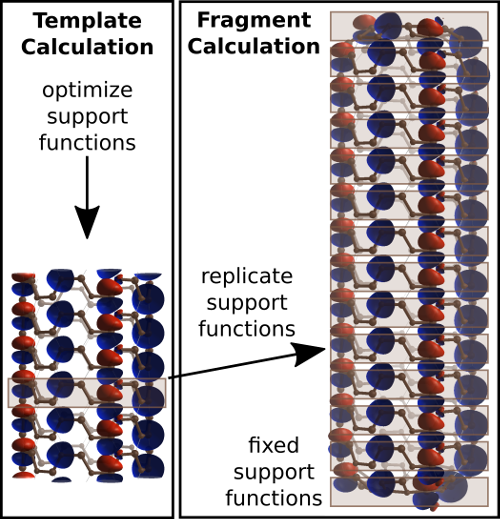}\label{fig:sic_schem_frp1}}
\subfigure[Four PFrag types generated in a four unit finite NT template.]{\includegraphics[scale=0.22]{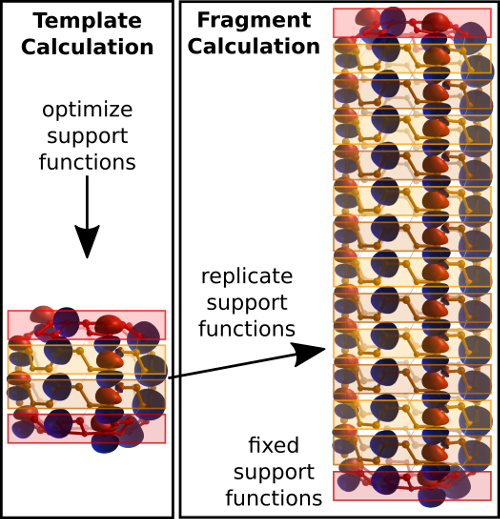}\label{fig:sic_schem_fr4}}
\subfigure[Six PFrag types generated in a six unit finite NT template.]{\includegraphics[scale=0.22]{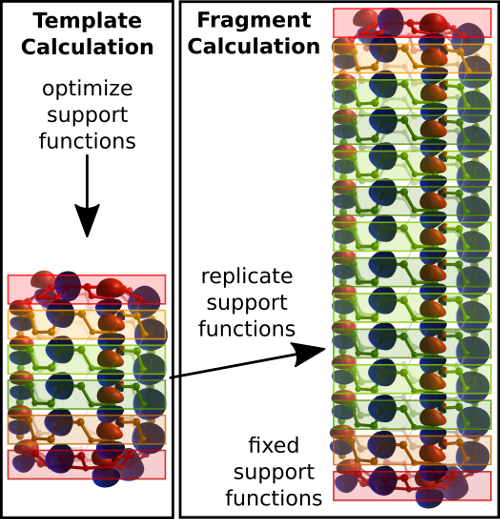}\label{fig:sic_schem_fr6}}
\subfigure[Eight PFrag types generated in an eight unit finite NT template.]{\includegraphics[scale=0.22]{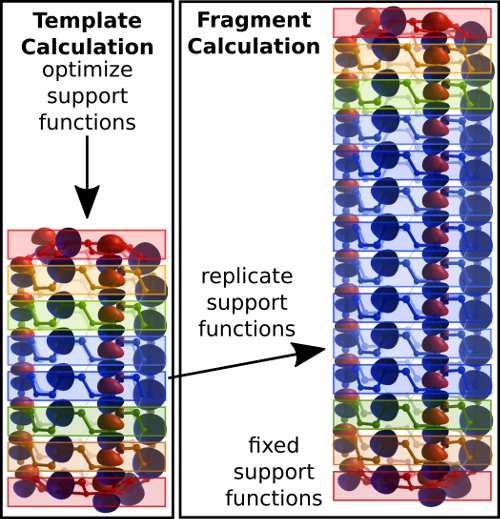}\label{fig:sic_schem_fr8}}
\subfigure[Seven PFrag types generated from a combination of a six unit periodic NT template and an eight unit finite NT template.]{\includegraphics[scale=0.22]{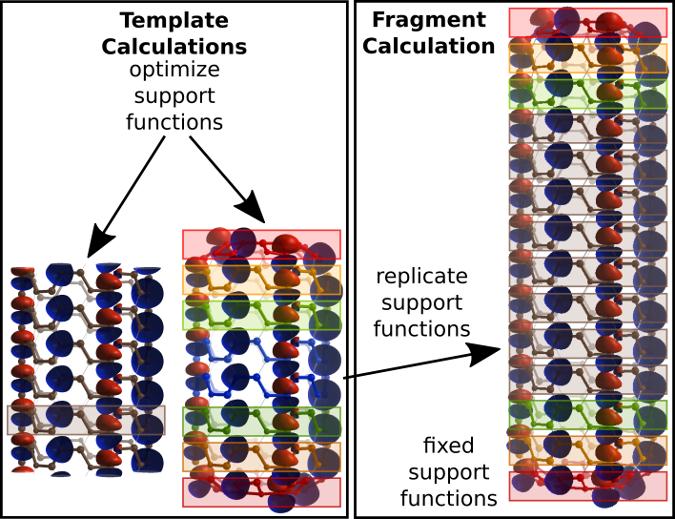}\label{fig:sic_schem_fr7}}
\caption{Schematic illustrating select PFrag setups for the finite SiCNT depicted in Fig.~\ref{fig:sic_posinp_free}.
The different colours indicate different PFrag types, with bonds within (between) pseudo-fragments
depicted in colour (grey).  Si atoms are depicted as larger spheres.
In each case the full length SiCNT of interest contains 14 units (392 atoms).}
\end{figure*}

\subsection{Edge Effects in a Finite SiC Nanotube}\label{sec:finite_sicnt}

We now consider what happens in the case of a finite SiCNT, i.e.\ what effect do the edges of
the NT for a given surface reconstruction have on the SFs?
We again take a 14 unit SiCNT, this time with the edge termination from Ref.~\cite{Pochet2010},
as depicted in Fig.~\ref{fig:sic_posinp_free}.
We expect that SFs associated with atoms near the centre of the SiCNT should be similar,
while those near the edges should differ.
As with the periodic NT, we start by calculating $S^{\textrm{onsite}}$,
which is plotted in Fig.~\ref{fig:sic_oo_free}, with selected SFs depicted in Fig.~\ref{fig:sic_tmb_free}.
As expected, the differences between edge SFs and central SFs are significant,
while the values of the $S^{\textrm{onsite}}$ vary smoothly as one moves along the NT.
Indeed, those in the centre, beyond around 10~\AA\ from the edge, show the same level of consistency as the SFs
of the periodic calculation, confirming that one can define a bulk-like central region.

The simplest option for performing a PFrag calculation would be to take the SFs
from the periodic template, as depicted in Fig.~\ref{fig:sic_schem_frp1}.
However, while this might represent a reasonable approximation for the centre of the NT,
this would clearly be unsatisfactory for the edges where the SFs are very different.
A better approach would therefore consist of defining several PFrag types, including edge,
bulk and, if needed, a (yet to be determined) number of intermediate types.
Following such intuition, confirmed by our values of $S^\textrm{onsite}$,
one could define a series of setups,
depicted in Figs.~\ref{fig:sic_schem_fr4}-~\ref{fig:sic_schem_fr8},
wherein the SFs are generated for a SiCNT which has an increasing number of intermediate PFrag types
and is therefore increasingly long.
We note that, as with the full system, we relaxed the geometries of the template structures for all PFrag setups using the parameters described in~\ref{app:comp}.

In the following we test each of the proposed PFrag setups, again comparing with the results of the full LS approach.
Although it is possible to account for reflections using the fragment approach implemented in BigDFT~\cite{Ratcliff2015a},
we choose to avoid doing so in order to simplify the calculation setup and analysis.
Therefore, taking the setup of Fig.~\ref{fig:sic_schem_fr4} as an example, this corresponds to
two separate edge PFrag types and two separate bulk PFrag types.  We therefore refer to this setup as 4 PFrags from now on.

Given the values of $S^{\mathrm{onsite}}$, we expect that SFs centred on atoms beyond around 10~\AA\
from the edge should be far enough away to be treated as bulk like.
Therefore, we expect the setup of~\ref{fig:sic_schem_fr8} to be sufficient to reproduce the correct electronic structure of the
LS reference calculation.
In order to determine whether or not this is indeed the case, we also take one additional setup, i.e.\ 10 PFrag types in
total.
Finally, we also consider an example of combining PFrags from the periodic and finite setups,
by replacing the central PFrag SFs of the 8 PFrag setup (Fig.~\ref{fig:sic_schem_fr8}) with those taken from
the periodic SiCNT.  This results in a setup consisting of 7 PFrags, as depicted in Fig.~\ref{fig:sic_schem_fr7}.

\begin{figure}
\centering
\subfigure[Densities of states.]{\includegraphics[scale=0.3]{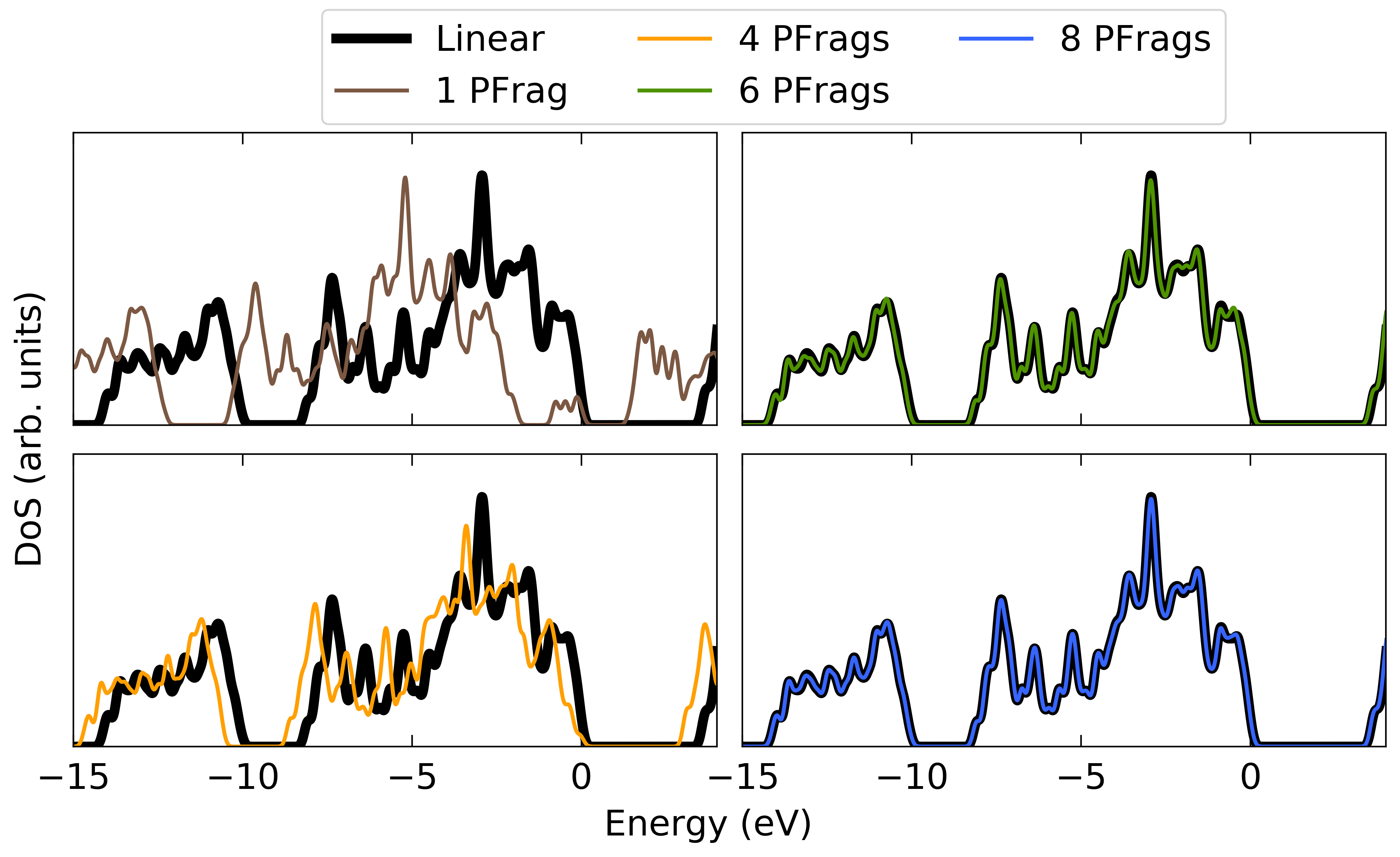}\label{fig:sic_free_dos}}
\subfigure[Averaged electron densities.]{\includegraphics[scale=0.3]{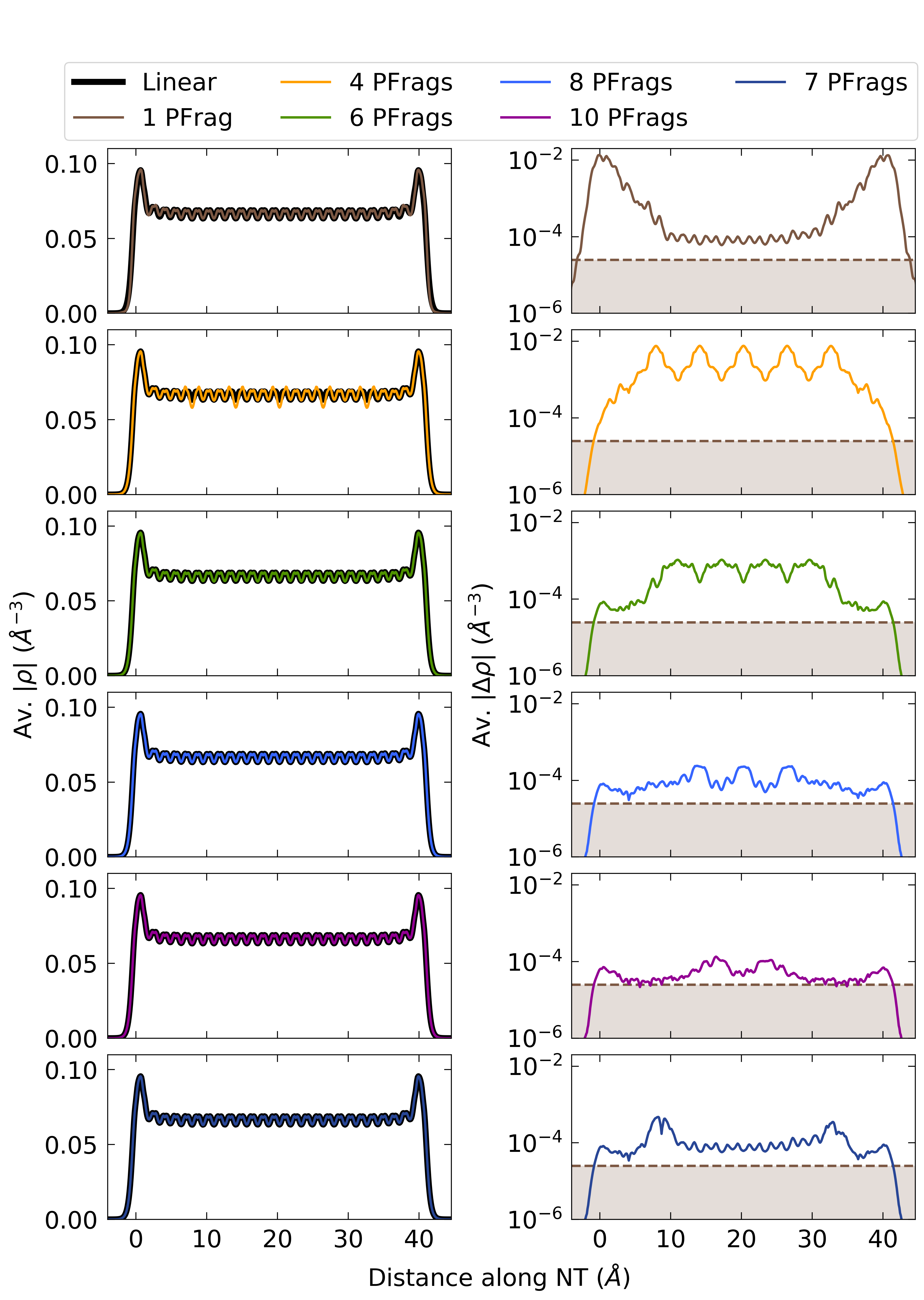}\label{fig:sic_free_rho}}
\caption{Panel \ref{fig:sic_free_dos}: DoS plots for different PFrag setups compared
to the LS reference.
The curves for 7 and 10 PFrags are indistinguishable from that for 8 PFrags, so the corresponding plots have been omitted.
Panel \ref{fig:sic_free_rho}: electron densities ($\rho$) averaged over the $xy$-plane [left]
and their corresponding errors ($\Delta \rho$) relative
to the LS reference [right] for different setups. The dashed line corresponds to the maximum average difference between the PFrag and LS approaches for the periodic NT.}
\end{figure}

The calculated energies and band gaps are given in Table~\ref{tab:sic_energies}, with the corresponding DoS and averaged electronic densities plotted in Figs.~\ref{fig:sic_free_dos}
and~\ref{fig:sic_free_rho} respectively.
As expected, the error for the 1 PFrag setup is very large.
Referring to the electronic density close to the edges, it is clear that the density is very poorly represented, as expected given that the PFrags
are all bulk-like.
This unsuitability of the SF basis is also reflected in the DoS, where the differences with the LS DoS
are significant.  As we increase the number of PFrag types, the error in the energy steadily decreases.
Indeed, the DoS converges very quickly with the number of PFrag types.

The band gap (compared to the LS reference) and electronic density are already well represented for
8 PFrag types, while the energy does not significantly change going to 10 PFrags.
These tests show that the $S^{\mathrm{onsite}}$ values are of great utility in
understanding the extensions of a defective region -- in this case the edges of the SiCNT.

Indeed, one may go a step further and quantitatively assess the relationship between $S^{\mathrm{onsite}}$ and the error in energy with respect to the LS calculation, $\Delta E$, as shown in Fig.~\ref{fig:sic_oo_v_e}.  As a simple test, we take the average value of $S^{\mathrm{onsite}}$ for a given PFrag type for the full SiCNT, e.g.\ for the 10 PFrag case this is the average across all SFs which are in the central bulk-like region.  This is compared to the corresponding $\Delta E$.
We also include the equivalent values for the periodic SiCNT on the same plot.
By fitting a linear function to the data, we show in Fig.~\ref{fig:sic_oo_v_e} that the average $S^{\mathrm{onsite}}$  at a given distance from the edge indeed provides a reliable indication of whether or not a given region may be treated as bulk-like or if additional PFrag types are required.

\section{Pseudo-Fragments and Lattice Distortions: the Cost Function $J$ for Error Estimation}\label{sec:wahba}

Thus far we have demonstrated the utility of the onsite overlap matrix for predicting whether a particular PFrag setup is likely to be appropriate for a given target system, for example by determining the lengthscale over which a perturbation might affect the form of the SFs.  In this section we now consider the related question of the suitability of a PFrag setup wherein the geometry of the template PFrags does not match that of the target system, i.e.\ the value of the cost function $J$ is non-zero.

As with the molecular fragment approach, the embedded PFrag method allows for small deviations in geometry between
template and system PFrags, however if these differences become large then additional types of PFrag may be required.
For systems with high levels of disorder and therefore very many PFrag types (i.e.\ little repetition) such an approach might therefore not be appropriate.
Stated otherwise, the $S^\mathrm{onsite}$ indicator might be useful only when there is not too much geometrical difference -- quantified by the cost function $J$ -- between the template PFrags and the target system.
It would therefore be useful to predict whether or not a PFrag setup is valid based on how much noise is present in the system.
That is, for a given value of the cost function $J$, what is the magnitude of the error introduced?

In order to answer this question, we first take the case of the pristine SiCNT and investigate the effect of adding random noise to the system.  We then return to the fixed length finite SiCNT discussed above, before finally considering finite SiCNTs of increasing length.

\begin{figure*}
\centering
\includegraphics[scale=0.3]{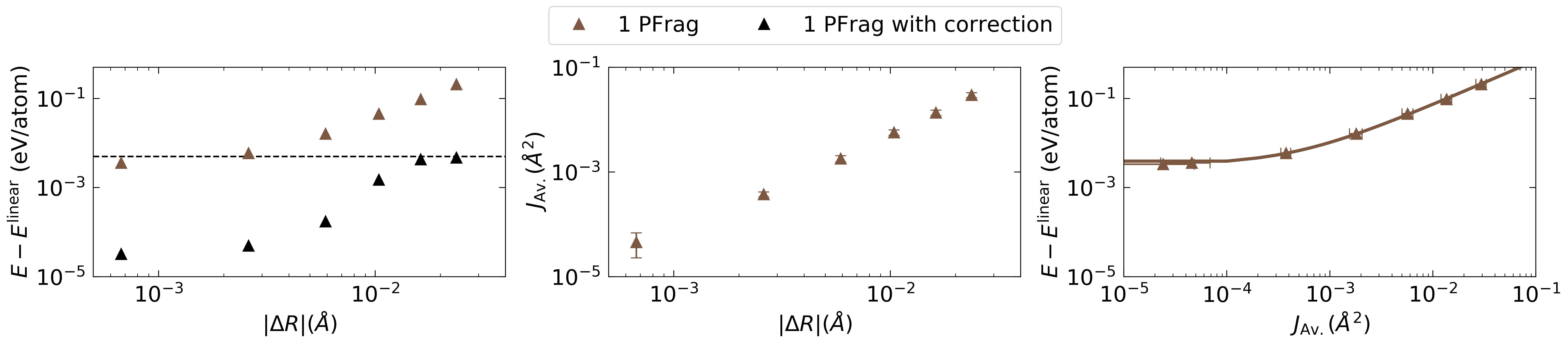}
\caption{Various quantities calculated for PFrag SFs generated for a pristine periodic SiCNT and used for a noisy SiCNT (`1~PFrag').  
Left panel: errors in total energy relative to fully optimized SFs ($E - E^{\mathrm{linear}}$) \emph{vs.} average difference between the noisy and pristine geometries ($|\Delta R|$).  The dashed line represents an absolute error of less than 5 meV/atom.
Centre panel: the average values of the cost function $J_{\mathrm{Av.}}$ \emph{vs.}\ $|\Delta R|$, where the error bars show the standard deviation in $J$.
Right panel: correlation between $J_{\mathrm{Av.}}$ and error in energy.  The linear fit has also been used to apply a correction to the energies, with the resulting errors shown on the left panel (`1 PFrag with correction').}\label{fig:sic_noise}
\end{figure*}

\subsection{Adding Noise: Effect of Imperfect Templates}\label{sec:noise}

To better understand if the cost function $J$ might be used as a quality indicator, we add increasing levels of random noise to the atomic positions of the 14 unit periodic SiCNT of Fig.~\ref{fig:sic_posinp_per}.
For each level of noise, we compare the total energy using fully optimized SFs with that using the PFrag SFs from the pristine NT.  For small levels of noise the PFrag setup should remain reasonable, while for high levels of noise the errors are expected to be significant.

For systems which contain more than one PFrag type, the cost function $J$ represents a useful quantity for describing the departure from rigid rototranslations.  In this case, each PFrag instance will have its own $J$ thanks to the added noise; in the following we take the average value, $J_{\mathrm{Av.}}$.
The level of noise for a given geometry may also be characterized by the average absolute difference between the noisy and pristine geometries, $\Delta R$, which in turn should be proportional to $J_{\mathrm{Av.}}$.
In the middle panel of Fig.~\ref{fig:sic_noise} we plot $\Delta R$ \emph{vs.}\ $J_{\mathrm{Av.}}$, showing that this is indeed the case.
In the left hand panel we show the error in the energy associated with the PFrag approach -- as expected this increases for higher levels of noise.
For $\Delta R \simeq 0.01$~\AA, the error is of the order of 50~meV/atom.

More interestingly, one can also assess the correlation between $J_{\mathrm{Av.}}$ and $\Delta E$.  As shown in the right hand panel of Fig.~\ref{fig:sic_noise}, these two quantities are well correlated within the range of values considered, being well described by a linear fit.  One could therefore imagine applying a correction to the PFrag energy by predicting the error based on the linear fit.
If one applies such a correction, as shown in Fig.~\ref{fig:sic_noise}, this results in an error of less than 5~meV/atom for each case, even for relatively large levels of noise.  In other words, the remaining error is purely due to that of the interpolation when roto-translating fragments.

For systems with little noise  ($J \lesssim 0.005$~\AA$^2$)
one can therefore apply the PFrag approach without inducing significant additional errors due to non-rigid PFrags.  For larger values of $J$, one can estimate fairly well the induced error. 
Clearly, this estimation might only be provided \emph{a posteriori}, i.e.\ if the value of $E^{\mathrm{linear}}$ is known. Nonetheless its evaluation might help identify the correlation between the average deviation of the cost function and the energy error of the PFrag approach, as we will show below.
We also note that this estimate mainly serves to indicate the bias induced by the PFrag approach, and would not necessarily be useful in practice for systems with significant distortions.  For example, it would not be easy to apply a similar correction to the forces, and in any case the error in the forces for a given PFrag setup is higher than that in the energy.  

Finally we note that in cases where the cost function is large, the PFrag setup might still offer a better initial guess for the SFs.  In the case of the noisiest geometry considered, using such a guess results in a factor of two reduction in the computational cost compared to using an initial guess based on atomic orbitals. In comparison, directly using the PFrag approach is more than 17 times quicker than the full LS approach for the same calculation.

\subsection{Effect of Non-Zero Cost Function for Finite SiC Nanotube}

We now return to the finite SiCNT of in Fig.~\ref{fig:sic_posinp_free}.
As mentioned in Section~\ref{sec:finite_sicnt}, the geometries were relaxed for all NTs, both the templates and the 14 unit target system.
As such, we expect to see non-negligible values of $J$, particularly for very short template systems,
where the differences in geometry with respect to the longer NT will be significant.  As we can see from Table~\ref{tab:sic_energies}, this is indeed the case, with a large value of $J_{\mathrm{Av.}}$ for the 1~PFrag case, and decreasing values for increasing numbers of PFrags.

Given the above considerations, we might estimate the error associated with the deviation in geometry between the template and target system PFrags, by using the linear fit between $J_{\mathrm{Av.}}$ and $\Delta E$  for the noisy system.  In other terms, we assume that the error coming from the non-zero cost function is equivalent to that due to random distortions in the geometry.
The values for the predicted error for the finite SiCNT are given in Table~\ref{tab:sic_energies}.
For large values of $J$ the errors are significantly overestimated, which is unsurprising given
this is well outside the regime used for the fit.
For smaller values of $J$ the errors 
agree relatively well with the actual errors.
Importantly, comparing the 8 and 10 PFrag calculations the further improvement in the energy can indeed be
attributed to the decrease in the cost function.
Such a qualitative behaviour is also present in the mixed periodic/finite PFrag setup.

\subsection{Finite SiC Nanotubes of Increasing Length}

Finally, we also consider SiCNTs of increasing length
to better understand the transferability of a given set of PFrags to similar systems of diverse sizes.
It was previously suggested~\cite{Pochet2010} that the energy $E_{\mathrm{NT}}$ of a SiCNT of a given diameter with $n$ units can be described by
\begin{equation}\label{eq:sic_size}
E_{\mathrm{NT}}\left(n\right) = \left(n - 2 n_{\mathrm{edge}} \right)E_{\mathrm{bulk}} + 2E_{\mathrm{edge}}\;,
\end{equation}
where $n_{\mathrm{edge}}$ is the number of units which can be considered to form the `edge' region at one end of the NT (i.e.\ the termination), $E_{\mathrm{edge}}$ is the corresponding energy and $E_{\mathrm{bulk}}$ is the energy of a bulk-like unit.  A value of $n_{\mathrm{edge}}=3$ was previously used, which agrees well with the above conclusions concerning both the distance from the edge of the NT beyond which the SFs revert to the equivalent bulk-like form, and the length beyond which the deviations in atomic coordinates (indicated by the cost function $J$) are small.   In the following we vary the length of the finite SiCNT, in order to determine how well this model is obeyed.  For each length of SiCNT, we relax the structures, as described in~\ref{app:comp}.
We compare the results of fitting the above model for full LS calculations, and for PFrag calculations using the 8 and 10 PFrag setups.

\begin{figure*}
\centering
\includegraphics[scale=0.3]{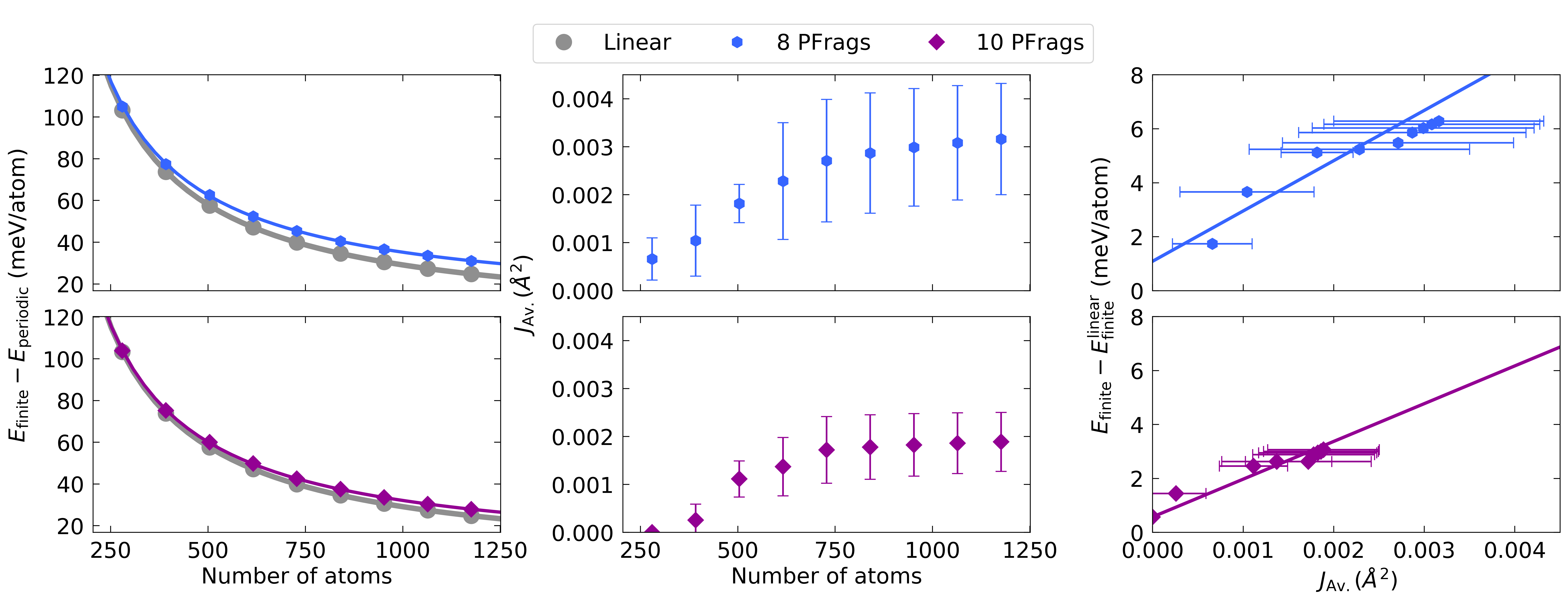}
\caption{Various quantities calculated for increasing length finite SiCNTs. Left panel: energy difference with respect to the periodic SiCNT for the LS and PFrag setups, where the line is a best fit to Eq.~\ref{eq:sic_size}.  Centre panel: average cost function values for increasing length SiCNTs, where the error bars denote the standard deviation in $J$.  Right panel: correlation between $J_{\mathrm{Av.}}$ and error in energy with respect to the LS reference.}\label{fig:sic_size}
\end{figure*}

The results are plotted in Fig.~\ref{fig:sic_size}.  For each scenario we have plotted both the actual energies obtained and the curve resulting from fitting the data to Eq.~\ref{eq:sic_size}.
It can be seen that for each setup the energies fit the model very well, confirming the above observations.  Furthermore, the agreement between the PFrag and LS setups is excellent, with only very small (less than 7~meV/atom) differences appearing for the longer SiCNTs.  Referring to the centre panel of Fig.~\ref{fig:sic_size}, it can be seen that this is due to larger values of $J$.  That is, the difference in geometry with respect to the template SiCNT increases with length until the point where relaxing the geometry has no further effects.  While the value of $J$, and correspondingly the associated errors, remain small, the cost function and the error in energy are again well correlated, as can be seen in the right panel of Fig.~\ref{fig:sic_size}.  It is therefore clear that $J$ can be seen as a general purpose indicator of the suitability of a given PFrag setup (i.e.\ the approximate induced errors) for cases where there are distortions in geometry, whether due to random noise as discussed above, or due to more systematic differences resulting from structural relaxations.

For the above calculations the PFrag approach is around eight times faster than the full LS approach.  The smaller speedup compared to the periodic NT calculations is due to the density taking longer to converge for the finite NT in the PFrag approach. Nonetheless, the computational savings are significant.

\section{Conclusion\label{sec:conclusion}}

The treatment of large systems in a density functional theory framework is a task which poses a number of challenges. The most important is the \emph{reduction} of the
number of degrees of freedom, within a paradigm with \emph{controllable accuracy}.
The linear scaling approaches such as those adopted in \textsc{onetep}, \textsc{Conquest} and \textsc{BigDFT} represent a first fundamental step in this direction: the support functions are
adapted to the chemical environment surrounding each region.
Such adaptivity allows the reduction of the degrees of freedom, with essentially no cost in terms of
accuracy, as the occupied Kohn-Sham orbitals are well represented \emph{by design} within such an approach.

This suggests that for large systems the degrees of freedom may be
reduced further, by identifying the various regions which might be expressed with the same
set of support functions, in other terms those regions which have a similar local chemical environment. This is important not only in view of computational savings,
but also to gain further insights into the systems' constituents. The reduction of the complexity of
the description is very important in this context:
performing a set of production quantum mechanical simulations with an approach which is unnecessarily costly would provide a study of poor quality,
as the simulation scheme would entangle interactions of different length scales and couplings.

A framework for identifying and exploiting regions with similar chemical environments has already proven very useful in the context of molecular systems,
where the search for systems' constituents naturally leads to fragmentation.
We have provided in this paper a similar approach for extended systems.  In such cases the concept of a separable fragment is not meaningful.  However, we have provided indicators which help to distinguish regions of an extended system which may be treated with identical degrees of freedom from regions which require a different set of support functions for an accurate simulation.

We have shown how a support function based pseudo-fragment approach may be applied to a SiC nanotube, demonstrating the applicability of such indicators in assessing the suitability of a given calculation setup.
This pseudo-fragment approach permits the exploitation of support function similarity in a whole new range of materials, and since there are no restrictions on cutting covalent bonds between pseudo-fragments, the user is free to define them according to their preference. 
In this work we focus on a quasi-one-dimensional system, however our approach might easily be applied to extended systems in two or three dimensions, for example for the treatment of defects.  Work in this direction is ongoing.
Both the setup and analysis of the presented calculations (including the generation of input files) has been done \emph{via} the use of a Jupyter Notebook, with the aim of easing both reproducibility of the results and the adoption by interested users.
We also note that in cases where a pseudo-fragment approach is not directly applicable, e.g.\ when there is significant distortion in the geometry of the system (as indicated by a cost function), one might still use it to generate a better input guess for the support functions.  This can significantly reduce the number of iterations required to further optimize the support functions.

The set of methods and indicators defined in this paper might also be considered as a first step towards
the control of the setup of more complex approaches, such as the treatment of defects with various levels of theory.
Currently we keep the support functions fixed, however this might be extended in future to allow the support functions to be further optimized in an active region and remain fixed in an environment region, i.e.\ to define an embedding method at the level of the basis.  
Long ranged interactions, which are not associated to a modification of the support functions, might therefore be treated with a different level of theory than short-ranged ones.
In such cases, the indicators presented in this work could be used to predict the size of the active region, thereby enabling an \textit{a priori} control of approaches that so far have primarily relied on the physical intuition of the user.

\section*{Acknowledgements}
LER acknowledges an EPSRC Early Career Research Fellowship (EP/P033253/1) and the Thomas Young Centre under grant number TYC-101.
This research used resources of the Argonne Leadership Computing Facility at
Argonne National Laboratory, which is supported by the Office of Science of the
U.S. Department of Energy under contract DE-AC02-06CH11357.
We are grateful to the UK Materials and Molecular Modelling Hub for computational resources, which is partially funded by EPSRC (EP/P020194/1).
This work also used the ARCHER UK National Supercomputing Service (http://www.archer.ac.uk).
LER thanks Damien Caliste for assistance with visualizing the support functions using v$\_$sim.

Calculation input and output files, and the Jupyter Notebook which has been used both to prepare the inputs and extract the data presented above, including the majority of the figures, are available at \url{https://gitlab.com/luigigenovese/pfrag-sicnt}.

\section*{References}

\bibliographystyle{iopart-num}
\bibliography{frag}

\appendix

\section{Computational details}\label{app:comp}

Calculations employed a grid spacing of $0.18$~\AA, corresponding to an accuracy of around $1$~meV/atom.  Four support functions were used for each atom, with localization radii of $4.5$~\AA, which were selected to give an error of $10$~meV/atom relative to the cubic scaling reference.    For the simulations in periodic boundary conditions, we used cell dimensions of 31.8 \AA\ in the $x$ and $y$ directions to reduce spurious interactions between periodic images.
For the LS calculations the FOE method was used to optimize the density kernel.  We used HGH pseudopotentials~\cite{Hartwigsen1998} and the PBE exchange correlation functional~\cite{Perdew1996}.  All calculations were performed at the $\Gamma$-point only.
Structures were relaxed using the cubic scaling approach using relatively loose convergence criteria, since the emphasis is on understanding the lengthscale over which an edge termination affects the electronic structure, rather than precisely converging the energies.  For the 14 unit structures and shorter template calculations the convergence criterion was a maximum force component below 0.03~eV/\AA, while for the longer finite SiCNTs a value of 0.1~eV/\AA\ was used.
For the density of states calculations, Gaussian smearing of 0.1~eV has been applied and the curves have been shifted so that the HOMO is at zero.

\section{Implementation Details}

\subsection{Multiple Instances of a Pseudo-Fragment}\label{app:mfi}

It should be noted that, unlike for an isolated template calculation, an embedded template calculation might include more than one instance of a given PFrag type. For example, in the case of the periodic SiCNT template calculation depicted in Fig.~\ref{fig:sic_schem_per} there are six instances of a single PFrag type.
In this case, the first instance of each PFrag type is the one for which the SFs and associated quantities are written to disk and thus reused for the full calculation, irrespective of whether this is `typical'.  For a template calculation where all instances of each PFrag type are truly identical, this point is irrelevant.  However in practice such PFrags will not always be identical, e.g.\ due to numerical noise for example coming from the egg-box effect or slightly differing chemical environments.

In principle, one could reduce the impact of arbitrarily selecting a single instance of a PFrag by performing an average over all instances, however the cost of performing this (including communicating the SFs) is non-negligible, and so it is instead left to the user to ensure the PFrag definitions are sensible, i.e.\ that the differences between different instances of the same PFrag are small.  To aid with this process, one could again refer to the onsite overlap matrix -- if, for example, in a template calculation, the onsite overlap between SFs on equivalent atoms on different instances of the same PFrag is much less than one, one can infer that the PFrags are distinct and an additional PFrag type should be defined.

\subsection{Initial Guess for the Density Kernel}\label{app:kernel}

For isolated fragment calculations, it is straightforward to build an initial guess for the KS coefficients (or equivalently, the density kernel) from the isolated coefficients (kernel).  However, the best procedure is less obvious for the embedded case, since it is not easy to separate the PFrag from its environment.
In particular, if one were to use very small PFrags which are strongly interacting with their environment (i.e.\ connected \emph{via} covalent bonds), it is not clear how one should extract the KS coefficients associated with a given PFrag -- it would be a rather poor approximation to cut the KS coefficients and retain only those associated with SFs belonging to that PFrag.  Similarly, taking and occupying only the lowest occupied KS orbitals and rejecting the higher energy occupied orbitals (in order to account for the lower number of electrons in the PFrag compared to the PFrag plus its environment) would result in a significant loss of information.

It is easier to cut the density kernel by retaining only matrix elements between SFs associated to the PFrag, however it still results in a loss of information.  In particular the total electronic charge is not correctly preserved.
For calculations with relatively large PFrags, this approach nonetheless results in a reasonable initial guess for the density kernel, while the total number of electrons reverts to the correct value after a few self-consistent iterations.

However, for calculations with much smaller PFrags, where the approximation is more severe, an alternative approach might be preferable.
In such cases, a more naive, conceptually more straightforward approach would be to generate a `diagonal' density kernel,  i.e.\ giving all SFs an equal weighting and normalizing to give the correct electron number.
This has the advantage of not introducing any particular bias, while also straightforwardly preserving the total number of electrons, and in practice has proven to be the most robust of the different options.

A final option is to randomly generate the density kernel (KS coefficients) and purify (orthonormalize).   However this approach is rather unstable compared to the alternatives.

\end{document}